\documentstyle[12pt,worldsci]{article}
\begin{document}
\def\n13{\hbox{$^{13}$N}}
\def\o15{\hbox{$^{15}$O}}
\title{HAS A STANDARD MODEL SOLUTION TO THE\\
 SOLAR NEUTRINO PROBLEM BEEN FOUND?}
\author{JOHN N. BAHCALL\footnote{\em Institute for
Advanced Study, School of Natural Sciences, Princeton,
NJ~08540}$^,$\footnote{\em Also, Institute for Nuclear Theory, University of
Washington, Seattle, WA 98195},
C. A. BARNES\footnote{\em W. K. Kellogg
Radiation Laboratory, Caltech, Pasadena, CA~91125},
J. CHRISTENSEN-DALSGAARD\footnote{\em Institut for Fysik og Astronomi,
Aarhus Universitet, DK-8000, Aarhus C, Denmark}, \\
B. T. CLEVELAND\footnote{\em Physics Department, University of
Pennsylvania, Philadelphia, PA 19104-6394},
S. DEGL'INNOCENTI\footnote{\em Departimento di Fisica dell'Universita di
Ferrara and Instituto Nazionale di Fisica Nucleare, Sezione di Ferrara,
I-44100, Ferrara, Italy}$^{,2}$,
B. W. FILIPPONE$^3$,\\
A. GLASNER\footnote{\em Racah
Institute of Physics, Hebrew University, Jerusalem, Israel and
Department of Astronomy and Astrophysics, University of Chicago,
Chicago, IL 60637}$^{,2}$,
R. W. KAVANAGH$^3$, S. E. KOONIN$^3$,
K. LANDE$^{5,2}$,\\
K. LANGANKE$^3$,
P. D. PARKER\footnote{\em Physics Department,
Yale University, New Haven, CT 06511}$^{,2}$,
M. H. PINSONNEAULT\footnote{\em Department of Astronomy,
Ohio State University, Columbus, OH 43210}$^{,2}$, \\
C. R. PROFFITT\footnotetext[0]{\em Computer Sciences Corporation,
IUE Observatory, Code 684.9,
Goddard Space Flight Center, Greenbelt, MD 20771}$^{10,2}$, and T.
SHOPPA$^3$
}
\maketitle

\abstract
The claim by Dar and Shaviv that they have found a standard model
solution to the solar neutrino problem is based upon an incorrect
assumption made in extrapolating nuclear cross sections and the
selective use of a small fraction of the nuclear physics and of the
neutrino data.  In addition, five different solar model codes show that
the rate obtained for the chlorine experiment using
the Dar-Shaviv stated parameters
differs by at least $14 \sigma$ from the observed rate.
\endabstract

\section{Introduction}
\label{sec-introduction}

In a widely circulated preprint, Dar and Shaviv\cite{dar94} claim
to have provided a standard solar model solution to the solar neutrino
problem. Their preprint has
aroused sufficient curiosity that we feel compelled to point out its
most obvious defects, although we do not plan  to publish our response
in a refereed journal unless the Dar and Shaviv
preprint is also published in a refereed journal.

To summarize our response,
Dar and Shaviv {\it have~not} solved the solar neutrino problem.
They have made an incorrect assumption  in extrapolating nuclear
cross sections
and have   used nuclear and neutrino data selectively.
Most surprisingly, their solar model results are not reproducible.
We obtain from
five independent solar neutrino codes and the Dar-Shaviv input
parameters, neutrino fluxes that are inconsistent with the
values reported by Dar and Shaviv.
With the Dar and Shaviv input parameters,
all four solar neutrino experiments differ from the solar model results
(obtained with the five different codes) by
more than $14 \sigma$ for the chlorine experiment $\bigg[\Big({\rm Rate
{}~(theory)} - {\rm Rate~(exp)}\Big)/\sigma ({\rm exp})\bigg]$\cite{davis64},
more than $ 4 \sigma$ for the Kamiokande experiment\cite{hirata91},
more than $4 \sigma$ for the GALLEX (gallium) experiment\cite{ansel93}, and
more than $3 \sigma$ for the SAGE (gallium) experiment\cite{abazov91}.

This response is organized as follows.
In \S 1, we comment on the fact that Dar and Shaviv chose to compare
their results with only a small part of the available chlorine
solar-neutrino data.  In \S 2, we point out that they made an {\it ad
hoc} and incorrect assumption in extrapolating the nuclear physics data
and compared their results
with only a small fraction of the available experiments.
In \S 3, we show that their solar model calculations are in disagreement
with calculations made using five other well-tested codes.

\section{Choosing Part of the Chlorine Solar-Neutrino Experimental Data}
\label{sec-choosing}

Dar and Shaviv chose to
consider (see their Figure~1) only four years of
data from the chlorine solar neutrino experiment beginning in 1987,
although 23 years of data have been reported$^2$. For the
period beginning in 1987, the experimental measurement of the chlorine
rate is $2.8 \pm 0.3$ SNU.  The measured rate for the entire
period for which data has been reported
is $2.28 \pm 0.23$ SNU.
The measured
rate in the chlorine experiment, during the short period
considered by Dar and Shaviv, is still more than $4 \sigma$ less than their
calculated result of $4.2$ SNU and is more than $14 \sigma$ from the
value we calculate (see \S 3 and Table~1 below)
for their stated parameters.

\section{Extrapolating Nuclear Cross Sections with an ad hoc Assumption and
Using Only Part of the Experimental Data}
\label{sec-extrapolating}

Dar and Shaviv claim that by replacing the conventional\cite{fowler84}
representation of the extra energy dependence
(in addition to the point-nuclei barrier penetration factor)
of the nuclear cross section factor
by another smoothly-varying, parameterized function,
they obtain a very different extrapolation to low energies for the
rate of the $^3He(\alpha,\gamma)^7Be$ reaction.
Of course, both procedures must yield the same answer if they
are both applied in an unbiased way to the same experimental data set
and all the energy dependences are included (see below).
For the reactions discussed by Dar and Shaviv, the existing data sets
determine well the extrapolations.

Dar and Shaviv, however, assumed incorrectly that the only
energy-dependent effect
besides point-nuclei barrier penetration
is nuclear size, and then compared their results with only
two of the nine published
experiments.  Their selective use of data and this incorrect
assumption explain why the Dar and Shaviv
answer for $S(0)$ differs from the standard value
obtained by nuclear physicists.  We explain in more detail below.

For extrapolation of nuclear cross section data to stellar energies,
it is conventional to define the astrophysical S-factor,
\begin{equation}
S(E) = \sigma(E)~ E~ {\rm exp} \{ 2 \pi \eta (E) \}.
\end{equation}
The Sommerfeld parameter is given by
\begin{equation}
\eta(E) = \frac{Z_1 Z_2 e^2}{\hbar v}\ ,
\end{equation}
where $v$ is the relative velocity of the two particles in the entrance
channel, $E$ is their relative energy,
and $Z_1,Z_2$ are their charges. The form of Eq. (1) accounts explicitly
for the energy dependence of s-wave tunneling
through the Coulomb barrier of two pointlike particles and
a kinematic flux factor. In the absence of
near-threshold resonances, the energy dependence of the S-factor is expected to
be small at low energies.
However, in order to carry out a reliable extrapolation,
the energy-dependent effects that are not accounted for in Eq. (1) must be
allowed to show up in $S(E)$. These effects include
nuclear structure, the
strong interaction, energy dependent operators in the transition matrix
elements, antisymmetrization between the colliding nucleons,
finite nuclear size, the final-state phase space, and the
contributions from other partial
waves. In fact, for all the reactions of
importance for the solar $p-p$ chain, the observed energy dependence of the
various reactions agrees well with that calculated in theoretical models
which account explicitly for the known nuclear effects
that are omitted from Eq. (1).  In particular,
it was demonstrated more than thirty years ago\cite{griffiths63}
that the complete Coulomb wave function provides a good description of
the measured cross section factor for the
$^2H(p,\gamma)^3He$ reaction
down almost to solar thermal energies (16 keV).

Dar and Shaviv have chosen to factor out a slightly different energy dependence
from the cross section data by defining a modified S-factor, ${\bar
S(E)}$, as
\begin{equation}
{\bar S}(E) =
\sigma(E) E {\rm exp} \{ 2 \pi \eta (E) [1 +2/\pi
(2 \sqrt{x} - {\rm arcsin} \sqrt{x} -\sqrt{x(1-x)} ) ) \},
\end{equation}
with $x=E/E_c$ and the Coulomb energy $E_c={Z_1 Z_2 e^2}/{R}$ at a
radius parameter $R$. With this definition, ${\bar S}(E)$ attempts to
account
for the finite size of the nuclei. However, ${\bar S}(E)$ is still expected to
be dependent on $E$, because of the other effects listed above
that introduce an energy
dependence in addition to the finite size of the nucleus.
The theoretical
models that have been used previously\cite{johnson92,kajino84,parker91}
to extrapolate the
cross section data to solar energies properly take account of
the finite  nuclear size effects
along with the other effects discussed above. Obviously both
definitions, Eqs. (1) and (3), must lead to the same results for the
low energy S-factor when proper account is taken of the additional energy
dependence not included explicitly in the respective equations.

Dar and Shaviv did not take account of the additional energy
dependences nor of all of the available nuclear physics data.
For the determination of the
low-energy cross section factor, S$_{34}$, for the $^3He(\alpha,\gamma)^7Be$
reaction,
Dar and Shaviv apparently adjusted the radius
parameter $R$ (see Eq. 3 above) so
that the energy dependence of ${\bar S}(E)$ is mostly removed for
two of the nine\cite{parker91} existing experiments.
(They seem not to have noticed that the value of $R$ that they obtain is
very different from the measured radius of $2.8$ fm determined by
electron scattering.)
They did not allow for the other energy dependences discussed above
and they only took
account of two of the experiments.

For the determination of the cross-section factor for the
$^7Be(p,\gamma)^8B$ reaction,
S$_{17}$, Dar and Shaviv first remove the contribution from d-wave
capture (based on previous calculations of this single component of the
energy dependence). However, the d-wave capture
contributes
about 6\% to the S-factor at the most effective energy for the
reaction (about 20 keV) and must be added to the computed s-capture rate.
In addition, previous
calculations\cite{johnson92,williams81}
demonstrate that there are significant further
contributions to the energy dependence for S$_{17}$ that must be included
in order to make
a reliable extrapolation. When these additional effects are included
in the extrapolation of ${\bar S}_{17}(E)$
(Eq. 3 above), the results must agree with the standard
extrapolations\cite{johnson92}.

Dar and Shaviv cited
the preliminary Coulomb
dissociation work described in preprint form\cite{moto94}
as evidence for a lower-than-standard value for the crucial
cross section factor for the $^7Be(p,\gamma)^8B$ reaction.
When the $E2$
contribution to this reaction is taken into account\cite{lang94},
the preliminary Coulomb-dissociation value differs from
the six direct measurements\cite{parker91} of the $^7Be(p,\gamma)^8B$
cross section
by a factor of two while the estimated uncertainty\cite{johnson92} in
direct measurements is only 11\%.
Moreover, there are still some
unanswered questions about the application of the
Coulomb-dissociation method for determining radiative capture cross
sections, aside from the experimental difficulties inherent in covering a
sufficient range in energy and angle to validate the reliability of
any inferences.

For other nuclear reactions,
Dar and Shaviv have used the low-energy cross section factors from an
earlier review\cite{caughlan88} which provided fitting
formulae suitable for use at temperatures ($\sim 10^9$ K)
much higher than are reached in
the sun ($\sim 10^7$ K).  The quantitative effect of these approximations is
difficult to estimate, especially since other authors (see section 4)
use--for solar
calculations--explicit formulae that are suitable for the lower solar
temperatures.  However, an approximate discussion of using the fitting
formulae for higher temperatures
at solar temperatures has been given\cite{bahcall92};
the principal effects of the high-temperature formulae are in
the direction to decrease the
predicted $^7$Be and $^8$B neutrino fluxes.

Dar and Shaviv discuss at some length the fact
that the Debye-H\"uckel approximation to the screened nuclear
potential is not correct everywhere in the sun.  It is not clear what they
recommend (although they say the effect is small),
nor if they are aware that modern
screening calculations go well beyond what they
discuss\cite{Graboske73,Carraro88}.

\section{Solar Model Calculations}
\label{sec-incorrect}
Six authors of this paper
(Bahcall, Christensen-Dalsgaard, Degl'Innocenti, Glasner, Pinsonneault,
and Proffitt)
have repeated the solar model calculations of Dar and Shaviv using
the non-standard parameters that Dar and Shaviv chose, namely, a solar
luminosity of $3.826 \times 10^{33}~{\rm erg~s^{-1} }$ and low energy
cross-section factors of $S_{34}(0)~=~0.45$ kev-b and $S_{17}(0)~=~17$
eV-b.
Dar and Shaviv did not specify in their preprint
many of the important input quantities in their model;
they did not state what they used for
the element abundances,
the radiative opacities, the equation of state,
and the neutrino cross sections.  They did not say which of the several
available perscriptions for diffusion they used.
We have therefore carried out calculations using a variety of different
choices for these quantities, namely, the choices made previously
as their best estimates by the six
different authors who used five independent stellar evolution
codes.\cite{bahpin94,proffitt94,Castellani93,Christensen-D93,bahglas94}

The results are shown in Table~1.
The first column identifies the computer code used in constructing the
solar model; the second column
indicates whether or not particle diffusion was included.  The
third, fourth and fifth columns give the calculated rates predicted for
the chlorine, gallium, and Kamiokande solar neutrino
experiments, respectively.
The numbers in parentheses indicate the number of standard
deviations quoted by the experimentalists (adding statistical and
systematic errors quadratically)
by which the calculated rates differ from the measured rates.
For gallium, we have compared with the most recent GALLEX
determination (which has smaller quoted uncertainties).
The last two columns of Table~1
give the fluxes of \n13 and \o15 solar neutrinos.
Some details regarding the solar models are given in the footnotes to
the table.

The calculations reported by Dar and Shaviv, the first row of Table~1,
predict a rate that differs by $8 \sigma$ from the
average chlorine experimental value.
Even if their calculations and assumptions were correct (and we think
they are not), then they {\it have not} solved the solar neutrino
problem unless one completely disregards the chlorine solar neutrino
experiment.

In rows two through four we give results obtained with five different
computer codes that have been intercompared in the literature with other
standard models; all of these codes have previously been shown to
give the same results for the same
input parameters and to give results that
agree well with still other well-tested solar model codes.
For the Dar and Shaviv parameters, all of the five well-calibrated
solar codes give results
very close to each other, as they must, but they all differ
significantly from the Dar
and Shaviv results. The relatively small differences between
the five well-calibrated solar
models are primarily due to the treatment of particle diffusion
(see column two).
This is a significant physical
effect that is somewhat difficult to include in solar model
calculations, but which has been taken into account in the most recent and
advanced solar model computer
programs.

For the Dar and Shaviv parameters, the five well-calibrated models give
for the chlorine experiment between 5.6 SNU ($14 \sigma$ discrepancy)
and 6.3 SNU ($17 \sigma$ discrepancy) depending upon
whether or not particle diffusion is included.  This contrasts with the
value of 4.2 SNU reported by Dar and Shaviv.  All five of the models
predict about 125 SNU ($4 \sigma $ discrepancy) for gallium, which
contrasts with the Dar and Shaviv value of 109 SNU.  Finally, the $^8$B
flux in the five well-calibrated solar models is between $4 \sigma$ and
$6 \sigma$ from the Kamiokande result (depending upon whether or not
particle diffusion is included), while the Dar and Shaviv result agrees
almost exactly with the Kamiokande measurement.

Even more striking differences exist for the \n13 and \o15 neutrino
fluxes, for which the Dar and Shaviv values differ by an order
of magnitude from the values obtained in the other programs.
Their flux for the \n13 neutrinos is low by about a factor of 6 and
their \o15 flux is low by about a factor of 15.

None of the well-tested solar codes that we have used are able to
reproduce the Dar and Shaviv results for neutrino fluxes.

\section{Summary}
\label{summary}

Dar and Shaviv did not succeed in solving the solar neutrino problem
despite the introduction of an {\it ad hoc} assumption, the selective
treatment of data, and the incorrect extrapolation of some  nuclear cross
sections. Their failure to solve the problem is not surprising since
it has been demonstrated elsewhere \cite{bahcall93} that the
chlorine and the
Kamiokande experiments are inconsistent with one another if one assumes
(on the basis of standard electroweak theory)
that the shape of the $^8$B neutrino spectrum is the same in the
laboratory and in the sun.

Several of the authors (JNB, SD, AG, KL, PDP, MHP, and CRP) thank the
Institute for Nuclear Theory at the University of Washington for its
hospitality and the Department of Energy for partial support during the
completion of this work.

\vfill\eject
$$\vbox{\halign{#\hfil\tabskip=1em plus1em
minus1em&\hfil#\hfil&\hfil#\hfil&\hfil#\hfil&\hfil#\hfil&\hfil#\hfil
&\hfil#\hfil\tabskip=0pt\cr
\multispan7{\hfil Table 1\hfil}\cr
\noalign{\medskip}
\multispan7{\hfil Solar Model Results for Dar and Shaviv
Parameters\hfil}\cr
\noalign{\medskip\hrule\smallskip\hrule\medskip}
\hfil
Solar Code\hfil&Diffusion&$^{37}$Cl&$^{71}$Ga&$^8$B$^{\displaystyle\dagger}$
&$^{13}$N$^{^{\displaystyle *}}$&$^{15}$O$^{^{\displaystyle *}}$\cr
&&(SNU)&(SNU)\cr
\noalign{\medskip\hrule\medskip}
Dar and
Shaviv&Yes?&4.2(8$\,\sigma$)&109(2.6$\,\sigma$)&2.8(0$\,\sigma$)&0.7&0.2\cr
\noalign{\bigskip}
YALE$^{\rm\displaystyle a}$&Yes&6.3(17$\,\sigma$)
&125(4$\,\sigma$)&4.2(6$\,\sigma$)&5.9&5.1\cr
\noalign{\bigskip}
AARHUS$^{\rm\displaystyle
b}$&No&5.6(14$\,\sigma$)&122(4$\,\sigma$)&3.6(4$\,\sigma$)&5.4&5.4\cr
&Yes&6.1(16$\,\sigma$)&124(4$\,\sigma$)&4.0(5$\,\sigma$)&5.8&5.8\cr
\noalign{\bigskip}
Proffitt$^{\rm\displaystyle c}$&Yes&6.3(17$\,\sigma$)
&126(4$\,\sigma$)&4.2(6$\,\sigma$)
&6.3&5.5\cr
\noalign{\bigskip}
FRANEC$^{\rm\displaystyle
d}$&No&5.8(15$\,\sigma$)&123(4$\,\sigma$)&3.9(5$\,\sigma$)&4.9&4.2\cr
\noalign{\bigskip}
ASTRA$^{\rm\displaystyle e}$&No&5.7(15$\,\sigma$)
&123(4$\,\sigma$)&3.7(4$\,\sigma$)&4.2&3.5\cr
\noalign{\medskip\hrule\medskip}
\multispan7{$^{\displaystyle\dagger}$Units: $10^6~{\rm cm}^{-2}{\rm
s}^{-1}$; ${^{\displaystyle *}}$Units: $10^8~{\rm cm}^{-2}{\rm s}^{-1}$.
\hfill}\cr
\multispan7{$^{\rm\displaystyle a}$cf.~Ref \cite{bahpin94}. Includes
helium and heavy element diffusion, as well as other
improvements.\hfill}\cr
\multispan7{$^{\rm\displaystyle b}$cf.~Ref \cite{Christensen-D93}. Second
model includes helium diffusion.\hfill}\cr
\multispan7{$^{\rm\displaystyle c}$cf.~Ref \cite{proffitt94}. Includes helium
and heavy element diffusion.\hfill}\cr
\multispan7{$^{\rm\displaystyle d}$cf.~Ref \cite{Castellani93}. Does not
include diffusion.\hfill}\cr
\multispan7{$^{\rm\displaystyle e}$cf.~Ref \cite{bahglas94}. Analagous to
the Best model without diffusion of ref
\cite{bahcall92}.\hfill}\cr
}}$$

\begin{thebibliography}{}
\bibitem{dar94}A. Dar, and G. Shaviv, Phys. Rev. Lett., preprint
submitted (1994).
\bibitem{davis64}R. Davis Jr., in Frontiers of Neutrino
Astrophysics, ed. Y. Suzuki, and K. Nakamura (Tokyo: Universal Academy
Press, Inc., 1993), p. 47.
\bibitem{hirata91}K. S. Hirata, Phys. Rev. D {\bf 44}, 2241 (1991);
Y. Suzuki, in Frontiers of Neutrino
Astrophysics, ed. Y. Suzuki, and K. Nakamura (Tokyo: Universal Academy
Press, Inc., 1993), p. 47.
\bibitem{ansel93}P. Anselmann, Phys. Lett. B {\bf 314}, 445 (1993); P.
Anselmann, et al., Phys. Lett., submitted, February (1994).
\bibitem{abazov91}A. I. Abazov, Nucl. Phys. B (Proc. Supppl.) {\bf 19},
84 (1991).
\bibitem{fowler84}W. A. Fowler, Rev. Mod. Phys. {\bf 56}, 149 (1984).
\bibitem{griffiths63}G. M. Griffiths, M. Lal, and C. D. Scarfe,
Can. J. Phys. {\bf 41}, 724 (1963).
\bibitem{johnson92}C. W. Johnson, E. Kolbe, S. E. Koonin, and
K. Langanke, Astrophys. J. {\bf 392}, 320 (1992).
\bibitem{kajino84}T. Kajino and A. Arima, Phys. Rev. Lett. {\bf 52},
739 (1984).
\bibitem{parker91}P. D. Parker, and C. Rolfs, C. in The Solar
Interior and Atmosphere, ed. A. Cox, W. C. Livingston, and M. S. Matthews
(Tucson: University of Arizona, 1991), p. 31
\bibitem{williams81}R. D. Williams, and S. E. Koonin, Phys. Rev.
C {\bf 23}, 2773 (1981).
\bibitem{moto94}T. Motobayashi, et al., Phys. Rev. Lett.,
submitted (1994).
\bibitem{lang94}K. Langanke, and T. Shoppa (CalTech preprint), Phys. Rev.
C, in press (April, 1994).
\bibitem{caughlan88}G. R. Caughlan, and W. A. Fowler, At. Data Nucl. Data
Tables {\bf 40}, 283 (1988).
\bibitem{bahcall92}J. N. Bahcall, and M. H. Pinsonneault, Rev.
Mod. Phys. {\bf 64}, 885 (1992).
\bibitem{Graboske73}H. E. DeWitt, H. C. Graboske, and M. S. Cooper,
Astrophys. J. {\bf 181}, 439 (1973).
\bibitem{Carraro88}C. A. Carraro, A. Schafer, and S. E. Koonin,
Astrophys. J. {\bf 331}, 565 (1988).
\bibitem{bahpin94}J. N. Bahcall, and M. H. Pinsonneault, in
preparation (1994).
\bibitem{proffitt94}C. R. Proffitt, Astrophys. J. {\bf 425}, 849
(1994).
\bibitem{Castellani93}V. Castellani, S. Degl'Innocenti, and G.
Fiorentini, Astron. and Astrophys. {\bf 271}, 601 (1993).
\bibitem{Christensen-D93}J. Christensen-Dalsgaard, C. R. Proffitt, and
M. J. Thompson, Astrophys. J. {\bf 403}, L75 (1993).
\bibitem{bahglas94}J. N. Bahcall, and A. Glasner, Astrophys. J., submitted
(1994).
\bibitem{bahcall93}J. N. Bahcall, and H. A. Bethe, Phys. Rev. D
{\bf 47}, 1298 (1993).
\end{thebibliography}
\end{document}